\documentclass[aps,preprint]{revtex4}
\usepackage[dvips]{graphicx}
\usepackage{amssymb}
\begin{document}
\draft

\title{\bf Resonances and poles in isoscattering microwave networks and graphs}
\author{Micha{\l} {\L}awniczak,$^{1}$ Adam Sawicki,$^{2,3}$  Szymon Bauch,$^{1}$ \\Marek Ku\'s,$^{2}$
and Leszek Sirko$^{1}$}

\address{$^{1}$Institute of Physics, Polish Academy of Sciences, Aleja  Lotnik\'{o}w 32/46, 02-668 Warszawa, Poland\\
$^{2}$Center for Theoretical Physics, Polish Academy of Sciences, Aleja  Lotnik\'{o}w 32/46, 02-668 Warszawa, Poland\\
$^{3}$School of Mathematics, University of Bristol, University
Walk, Bristol BS8 1TW, UK}

\date{February 27, 2014}
\bigskip

\begin{abstract}
{\it Can one hear the shape of a graph?} This is a modification of the famous question of Mark Kac "Can one hear the shape of a drum?" which can be asked in the case of scattering systems such as  quantum graphs and microwave networks. It addresses an important mathematical problem whether scattering properties of such systems are uniquely connected to their shapes?

Recent experimental results based on a characteristics of graphs such as the cumulative phase of the determinant of the scattering matrices indicate a negative answer to this question (O. Hul, M.~{\L}awniczak, S. Bauch, A. Sawicki, M. Ku\'s, L.
    Sirko, Phys. Rev. Lett 109, 040402 (2012).).

In this paper we consider important local characteristics of graphs such as structures of resonances and poles of the determinant of the scattering matrices. Using these characteristics we study experimentally and theoretically properties of graphs and directly confirm that the pair of graphs considered in the cited paper is isoscattering. The experimental results are compared to the theoretical ones for a broad frequency range from 0.01 to 3 GHz. In the numerical calculations of the resonances of the graphs absorption present in the experimental networks is taken into account.

\end{abstract}

\pacs{03.65.Nk,05.45.Ac}

\bigskip
\maketitle

\section{Introduction. Isospectrality and isoscattering}

Let us start with a natural definition: two vibrating systems are isospectral if and only if their spectra are identical. The famous question posed by Marc Kac in 1966 "Can one hear the shape of a drum?" \cite{Kac66} addresses the problem whether two isospectral drums have the same shape. In mathematical terms it reduces to a question of uniqueness of spectra of the Laplace operator on the planar domain with Dirichlet boundary conditions.
The negative answer was given in 1992 when Gordon, Webb, and Wolpert \cite{Gordon92a,Gordon92b}, using Sunada's theorem \cite{Sunada85}, found a way to construct pairs different in shape but isospectral domains in $\mathbb{R}^{2}$. The procedure of designing such isospectral planar domains consists of appropriate cutting the 'drum' into subdomains and rearranging them into a new one with the same spectrum. An experimental confirmation that `hearing' the shape is impossible was presented by Sridhar and Kudrolli \cite{Sridhar1994} and Dhar et al. \cite{Dhar2003} for a pair of isospectral microwave cavities.

An analogous question can be posed not only for other mechanical vibrating systems but in all situations where time evolution can be analyzed in terms of eigenmodes and corresponding eigenfrequencies forming the spectrum of the system. The problem of isospectrality for quantum graphs was analyzed by Gutkin and Smilansky \cite{Gutkin01}. Such graphs consist of one-dimensional bonds connecting vertices. In each bond the wave propagation is governed by the one-dimensional Scgr\"odinger equation.  Gutkin and Smilansky proved that the spectrum identifies uniquely the graph if the lengths of its bonds are incommensurate. In the opposite situation of commensurate lengths of bonds there exist graphs with different metric and topological properties having equal spectra. A general method of construction of isospectral graphs \cite{Band09,PB09} uses the extended Sunada's approach. Here also one cuts the graph and "transplants" the pieces into a different arrangement.  As a result of such transplantation every eigenfunction of the first graph one can assign an eigenfunction of the second one with the same eigenvalue.

Inability of determining the shape from the spectrum alone does not preclude possibilities of distinguishing one drum from another in more sophisticated experiments. Indeed, basing on numerical simulations Okada et al.\ \cite{OSTH05} conjectured that isospectral domains constructed by Gordon, Webb and Wolpert can be in fact distinguished in scattering experiments by different distributions of poles of the scattering matrices. One can thus ask whether also the geometry of a graph can be determined in scattering experiments.

The question was answered negatively by Band, Sawicki and Smilansky
\cite{Sawicki10,Sawicki10b}. They analyzed isospectral quantum graphs with attached infinite leads and developed a method of constructing pairs of graphs which are called \textit{isoscattering}. By definition, isoscattering graphs are isopolar when their scattering matrices have the same poles or isophasal when the phases of the determinants of their scattering matrices are equal.

Isopolar lossless graphs need not be isophasal since to determine the phases one needs more information.
In contrary, any two isophasal lossless graphs are isopolar, as analytic continuations of the phases for complex wave vector $k$ overlap if the phases overlap for real $k$ values. Knowing phases we know determinants of the scattering matrices and therefore we know poles structures. Note however that knowing poles does not determine phase of the determinant completely. Therefore it is a priori possible to have two lossless graphs that are isopolar but not isophasal. The construction of such a pair cannot be addressed using the method described in \cite{Sawicki10} and requires a new approach which is beyond the scope of the paper. The situation is even more complicated in the case of graphs with internal absorption such as microwave networks as to determine determinants of the scattering matrices we need to know both the phases and amplitudes. In such cases theoretical models which inherently involve some simplifications should be confronted with the experimental results.

In \cite{Sawicki10,Sawicki10b} the authors showed that any pair of isospectral quantum graphs obtained by
the above outlined method of \cite{Band09,PB09} is isoscattering if the infinite leads are attached in a way preserving the symmetry of the isospectral construction \cite{Sawicki10,Sawicki10b}.

Finally, in an experiment by Hul et al.\ \cite{Hul12} with two isoscattering microwave networks simulating a pair of isoscattering quantum graphs the existence of topologically different graphs with the same isophasal features was experimentally confirmed.

\section{Microwave networks simulating quantum graphs}

Quantum graphs are idealizations of physical networks in the limit where the lengths of the wires are much larger than their diameter. A detailed theoretical analysis of their properties as well as applications in modeling various physical problems can be found in \cite{Gnutzmann2006} and references cited therein. Methods of their experimental realizations were presented in \cite{Samuelson2004,Heo2008}.

In 2004 Hul et al.\ \cite{Hul04} showed that quantum graphs could be successfully modeled by microwave networks. The introduction of one-dimensional microwave networks simulating quantum graphs extended substantially the number of systems, like highly excited hydrogen atoms \cite{Blumel1991,Bellerman1992,Sirko1993,Sirko1996,Sirko2002} and two-dimensional microwave  billiards  \cite{Stockmann90,Sridhar,Richter,Sirko97,Bauch98,Sirko2000,Blumel2001,Hlushchuk2001, Savytskyy2004,Hemmady2005,Kober2010,Barkhofen2013}, used to verify wave effects predicted on the basis of quantum physics. The later papers on this topic \cite{Hul2005,Lawniczak2008,Lawniczak2010} clearly demonstrated that microwave networks can be successfully used to investigate properties of quantum graphs, also with highly complicated topology and absorption.

A microwave network consists of $n$ vertices connected by $B$ bonds,   e.g., coaxial cables. The $n\times n$ connectivity matrix $C_{ij}$ of a network takes the value $1$ if the vertices $i$ and $j$ are connected and $0$ otherwise.  Each vertex $i$ of a network is connected to the other vertices by $v_{i}$ bonds, $v_{i}$ is called the valency of the vertex $i$.

A coaxial cable consists of an inner conductor of radius $r_1$ surrounded by a concentric conductor of inner radius $r_2$. The space between the inner and the outer conductors is filled with a homogeneous material having the dielectric constant $\varepsilon$. Below the onset of the next TE$_{11}$ mode \cite{Jones}, inside a coaxial cable can propagate only the fundamental TEM mode, in the literature called a Lecher wave.

Using the continuity equation for the charge and the current one can find the propagation of a Lecher wave inside the coaxial cable joining the $i$--th and the $j$--th vertex of the microwave network \cite{Landau,Hul04}. For an ideal lossless coaxial cable the procedure leads to the telegraph equation on the microwave network

\begin{equation}
\label{1}
\frac{d^2}{dx^2}U_{ij}(x)+\frac{\omega^2\varepsilon}{c^2}U_{ij}(x)=0,
\end{equation}

where  $U_{ij}(x,t)$  is the potential difference between the conductors, $\omega =2\pi\nu$  is the angular frequency and $\nu$ is the microwave frequency, $c$ stands here for the speed of light in a vacuum, and $\varepsilon$ is the dielectric constant.

Upon the correspondence: $\Psi_{ij}(x)\Leftrightarrow U_{ij}(x)$ and $k^2\Leftrightarrow \frac{\omega^2 \varepsilon}{c^2}$, the equation (\ref{1}) is formally equivalent to the one--dimensional Schr\"{o}dinger equation (with $\hbar=2m=1$) on the graph possessing time reversal symmetry \cite{Kottos}

\begin{equation}
\label{2}
\frac{d^2}{dx^2}\Psi_{ij}(x)+k^2\Psi_{ij}(x)=0.
\end{equation}

\section{Experimental setup}

In the following we consider the pair of microwave networks which was already described in details in \cite{Hul12}. In Fig.~1a and Fig.~1b we show  the two isoscattering graphs which are obtained from the two isospectral ones by attaching two infinite leads $L^{\infty}_1$ and $L^{\infty}_2$.  Using microwave coaxial cables we constructed the two microwave isoscattering networks shown in Fig.~1c and Fig.~1d. In order to preserve the same approximate size of the graphs in Fig.~1a and Fig.~1b and the networks in Fig.~1c and Fig.~1d, respectively, the lengths of the graphs were rescalled down to the physical lengths of the networks, which differ from the optical ones by the factor $\sqrt{\varepsilon}$, where $\varepsilon \simeq 2.08$ is the dielectric constant of a homogeneous material used in the coaxial cables.

For the above networks and graphs we will consider two most typical
physical vertex boundary conditions, the Neumann and Dirichlet ones. The first one imposes
the continuity of waves propagating in bonds meeting at $i$ and vanishing of the sum
of their derivatives calculated at
the vertex $i$. The latter demands
vanishing of the waves at the vertex.

The graph in Fig.~1a consists of $n=6$ vertices connected by
$B=5$ bonds. The valency of the vertices $1$ and $2$ including leads is
$v_{1,2}=4$ while for the other ones $v_i=1$. The vertices with numbers $1,2,3$ and $5$, satisfy
the Neumann vertex conditions, while for the vertices $4$ and $6$
we have the Dirichlet ones.
The second graph, shown in Fig.~1b, consists of $n=4$ vertices connected by $B=4$
bonds.  The vertices with
the numbers $1,2$ and $3$ satisfy the Neumann vertex conditions
while for the vertex $4$, the Dirichlet condition is imposed.

The bonds of the microwave networks have the following optical lengths:
\begin{tabular}{ l r }
\\
 $ a=0.0985\pm 0.0005\mbox{ m, }$ \\
 $ b=0.1847 \pm 0.0005\mbox{ m, }$ \\
 $ c=0.2420 \pm 0.0005 \mbox{ m, }$ \\
 $2a=0.1970\pm 0.0005\mbox{ m, }$ \\
 $2b=0.3694 \pm 0.0005\mbox{ m, }$ \\
 $2c=0.4840 \pm 0.0005 \mbox{ m. }$\\
\label{LConst}
\end{tabular}

The uncertainties in the bond lengths of the networks are connected  with the preparation of Dirichlet and Neumann $v_{1,2}=4$ vertices. The Dirichlet vertices were prepared by closing the cables with brass caps to which the internal and external conductors of the cables were soldered. In the case of the Neumann vertices the internal wires of the cables were soldered together.

The systems are described in terms of $2\times 2$ scattering matrix
$S(\nu)$
\begin{equation}
\label{3}
S(\nu)=\left( \matrix{S_{1,1}(\nu)&S_{1,2}(\nu)\cr
S_{2,1}(\nu)&S_{2,2}(\nu)} \right) \mbox{,}
\end{equation}
relating the amplitudes of the incoming and outgoing waves of frequency $\nu$ in both leads.

In order to measure the two-port scattering matrix $S(\nu)$ the vector network analyzer (VNA) Agilent E8364B was connected to the vertices $1$ and $2$ of the microwave networks shown in Fig.~1c and Fig.~1d. We performed the measurements of the scattering matrix $S(\nu)$ in the frequency range $\nu = 0.01-3$ GHz. The connection of the VNA to a microwave network (see Fig.~1e) is equivalent to attaching of two infinite leads to a quantum graph.

\section{Resonances of amplitudes and poles of scattering matrices}
Two networks in Fig.~1c and Fig.~1d are isopolar if their scattering matrices have the same poles.
In order to study isopolar properties of graphs presented in Fig.~1 we consider important local characteristics of graphs such as structures of experimentally measured resonances and theoretically evaluated poles of the determinant of the two-port scattering matrices.  Such a comprehensive analysis is important since for open systems resonances show up as poles \cite{Kottos2003,Borthwick2007} occurring at complex wave numbers $k_l=\frac{2\pi}{c}(\nu_l -i\Delta\nu_l)$, where $\nu_l$ and $2\Delta\nu_l$ are associated with the positions and the widths of resonances, respectively.
In Fig.~2a we show that for the frequency range from 0.01 to 3 GHz  the amplitudes $|\det\bigl(S^{(I)}(\nu)\bigr)|$ and $|\det\bigl(S^{(II)}(\nu)\bigr)|$ of the determinants of the scattering matrices $S^{(I)}(\nu)$ and $S^{(II)}(\nu)$ of the networks shown in Fig.~1c and Fig.~1d, respectively, are close to each other, clearly showing that we are dealing with the isoscattering networks.  The results obtained  for the networks presented in Fig.~1c and in Fig.~1d are marked by blue full squares and red open circles, respectively. For the lower frequency range the agreement between the results obtained for both networks are almost perfect. However, for the frequency range 2-3 GHz small discrepancies arise, both in the phases and amplitudes of the determinants of the scattering matrices. They are possibly caused by a small differentiation of the vertex boundary conditions of the networks in the function of frequency $\nu$. It was shown in the paper \cite{Hul12} that the modification of the boundary conditions may cause significant changes in the spectra of the networks.

The analytical formulas for the elements of the scattering matrices $S^{(I)}(k)$ and $S^{(II)}(k)$ are presented in the Appendix. Simple but extensive calculations showed that both scattering matrices posses the isoscattering properties. Therefore, in Fig.~2b using the contour plot we present only the poles of the amplitude of the determinant of the scattering matrix $|\det\bigl(S^{(II)}(k)\bigr)|$ (solid circles) calculated for the graph with $n=4$ vertices (Fig.~1b) for the frequency range from 0.01 to 3 GHz. The numerical calculations were performed for the isoscattering graph having the same bond lengths as the ones measured for the microwave network presented in Fig.~1d. We also imposed the proper vertex boundary conditions. The vertical axis of Fig.~2b shows the imaginary part $\Delta\nu$ of the poles of the graph. Fig.~2a and Fig.~2b  clearly show very good agreement between the positions of the experimental scattering resonances and the theoretical poles. To make this comparison even more straightforward the poles of the determinant of the scattering matrix $\det\bigl(S^{(II)}(k)\bigr)$  are marked in Fig.~2a by big solid circles.
Also the full widths at the half maximum $\Delta \nu_l^{exp}$ of the experimental resonances obtained by a numerical fitting of a single or a sum of Lorentz functions to a single or multiple resonances, respectively, are in good agreement with the theoretical ones $2 \Delta \nu_l$. The averaged ratio of the resonances widths $\langle \frac {\Delta \nu_l^{exp}}{2 \Delta \nu_l} \rangle_l = 0.99 \pm 0.13$.

The paper \cite{Hul04} shows that loses in the networks can be described by treating the wave number $k$  as a complex quantity with absorption-dependent imaginary part $\mathrm{Im}\Bigl[k\Bigr]=\beta \sqrt{2\pi \nu/c }$ and the real part $\mathrm{Re}\Bigl[k\Bigr] =  2\pi \nu/c $, where $\beta$ is the absorption coefficient and $c$ is the speed of light in vacuum. The analytical formulas for the theoretical scattering matrices $S^{(I)}(k)$ and $S^{(II)}(k)$ allow us to reconstruct the resonances in the amplitudes of the determinants of the scattering matrices $|\det\bigl(S^{(I)}(k)\bigr)|$ and $|\det\bigl(S^{(II)}(k)\bigr)|$. The solid line  in Fig.~2a shows the amplitude of the determinant of the scattering matrix $|\det\bigl(S^{(II)}(k)\bigr)|$ calculated for the absorption coefficient $\beta=0.00762 m^{-1/2}$. Fig.~2a shows that especially for the lower frequency range 0.01-2 GHz the theoretical results are in very good agreement with the experimental ones. Also for higher frequency range from 2 to 3 GHz the overall agreement with the experimental results is good. However, some departures from the experimental results are visible. As mentioned above they are very likely caused by small departures of the vertex boundary conditions of the networks for higher frequencies $\nu $ from the assumed in the calculations ideal Neumann and Dirichlet boundary conditions.

Our experimental data allow us also to check the isophasal properties of the studied networks for a broader frequency range 0.01-3 GHz than the previously reported in \cite{Hul12}, which was up to 1.7 GHz.

Two networks in Fig.~1c and Fig.~1d are isophasal if the phases
\begin{equation}
\label{4}
\mathrm{Im}\Bigl[\log\Bigl(\det\bigl(S^{(I)}(\nu)\bigr)\Bigr)\Bigr]=
\mathrm{Im}\Bigl[\log\Bigl(\det\bigl(S^{(II)}(\nu)\bigr)\Bigr)\Bigr]\mbox{,}
\end{equation}

of the determinants of the scattering matrices $S^{(I)}(\nu)$ and $S^{(II)}(\nu)$ of the networks shown in Fig.~1c
and Fig.~1d, respectively, are equal for all values of $\nu$.

 In Fig.~2c we show that also for higher frequency range from 0.01 to 3 GHz the cumulative phases of the determinants of the scattering matrices of the studied networks are close to each other, confirming that we are also dealing  with the isophasal networks.  The results obtained  for the networks presented in Fig.~1c and in Fig.~1d are marked in Fig.~2c by blue full squares and red open circles, respectively.

In summary, we investigated experimentally resonances of the two microwave networks which were constructed to be isoscattering. Linking the structures of experimental resonances with the local characteristics of graphs such as theoretically calculated poles of the determinant of the scattering matrices we  demonstrated that the networks are iospolar, i.e., isoscattering, within the experimental errors, for a broad microwave frequency range up to 3 GHz.  We also reconstructed theoretically the scattering resonances of the isoscattering microwave networks and compared them to the experimental findings. The results  clearly show very good agreement between the theoretical and the experimental scattering resonances. As an additional test of isoscattering of the networks we  demonstrated that the networks are isophasal for the same broad frequency range up to 3 GHz.  Finally, one should remark that our results clearly show that it is possible to construct the experimental networks which are isoscattering or nearly isoscattering for a broad frequency range. It opens the field for the possible future electronic applications.

This work was partially supported by the Ministry of Science and Higher Education grant No. N N202 130239.

\section{Appendix}

For details of calculations of scattering matrices for graphs one should consult \cite{Sawicki10,Sawicki10b} (see also \cite{Gnutzmann2006} for general principles of wave propagation in quantum graphs). Here we present a brief account of the necessary calculations.

For brevity of notation it is convenient to denote the wave propagating through an edge (bond)  $e$ by $\Psi_e$, i.e.\ use edges to index the waves rather than corresponding vertices as in Eq.~(\ref{2}). Propagation in each edge is described by the free Schr\"odinger equation
\begin{equation}\label{s1}
-\frac{d^2}{dx_e^2}\Psi_e(x_e)=k^2\Psi_e(x_e),
\end{equation}
where $x_e$ is a coordinate parameterizing the edge $e$. Mathematically speaking the propagation in the whole graph is dictated by the Laplace operator on the graph which is the sum of one-dimensional Laplacians, $-d^2/dx_e^2$, each acting on the corresponding edge.

The solution of (\ref{s1}) for each edge takes the form
\begin{equation}\label{psie}
\Psi_e(x_e)=a^{in}_e\exp(-ikx_e)+a^{out}_e\exp(ikx_e).
\end{equation}
In particular, for the two leads $L^{\infty}_1$ and $L^{\infty}_2$, we have
\begin{equation}\label{psil}
\Psi_l(x_l)=a^{in}_l\exp(-ikx_l)+a^{out}_l\exp(ikx_l), \quad l=1,2.
\end{equation}
If the wave with a wave number $k$ propagates in the whole graph, i.e.\ $k^2$ is an eigenvalue of the graph Laplacian (for a scattering graph the spectrum of eigenvalues is, in general, continuous), the solutions (\ref{psie}) and (\ref{psil}) satisfy the vertex boundary conditions for a particular graph. Imposing the conditions, we obtain a linear set of equations connecting the amplitudes $a^{in}_e$ and $a^{out}_e$ of the forward and backward propagating waves in the edges, as well as incoming and outgoing amplitudes $a^{in}_l$ and $a^{out}_l$ in the leads. The equations can be solved for $a^{out}_{1,2}$ in terms of $a^{in}_{1,2}$,
\begin{equation}\label{scatap}
\left(
\begin{array}{c}
 a_1^{out} \\
 a_2^{out} \\
\end{array}
\right)=
\left(
\begin{array}{cc}
 S_{1,1}(k) & S_{1,2}(k) \\
 S_{2,1}(k) & S_{2,2}(k) \\
\end{array}
\right)
\left(
\begin{array}{c}
 a_1^{in} \\
 a_2^{in} \\
\end{array}
\right).
\end{equation}
Applying this procedure to the graphs in Fig.~1a and Fig.~1b we get, respectively

\small
\[
S^{(I)}_{1,1}(k)=\]
\[\frac{\left(-1+e^{4 i b k}\right)
\left(-1+e^{4 i c k}\right)-2
\left(-1+e^{4 i (b+c) k}\right) \cos\left(2 a k\right)-2 i
\left(-1+e^{2 i (b+c) k}\right)^2 \sin\left(2 a k\right)}
{\left(-3+e^{4 i b k}+e^{4 i c k}+e^{4 i (b+c) k}\right)
\cos\left(2 a k\right)-i \left(-5+e^{4 i b k}+e^{4 i c k}+4 e^{2 i (b+c) k}-e^{4 i (b+c) k}\right)
\sin\left(2 a k\right)},
\]

\[
S^{(I)}_{1,2}(k)=\]
\[\frac{2 \left(-e^{2 i b k}-e^{2 i c k}+e^{2 i (2 b+c) k}+e^{2 i (b+2 c) k}\right)
\sin(2 a k)}
{i \left(-3+e^{4 i b k}+e^{4 i c k}+e^{4 i (b+c) k}\right)
\cos(2 a k)+\left(-5+e^{4 i b k}+e^{4 i c k}+4 e^{2 i (b+c) k}-e^{4 i (b+c) k}\right)
\sin(2 a k)},
\]

\[
S^{(I)}_{2,1}(k)=S^{(I)}_{1,2}(k),
\]

\[
S^{(I)}_{(2,2)}(k)=\]
\[\frac{-\left(-1+e^{4 i b k}\right) \left(-1+e^{4 i c k}\right)-2
\left(-1+e^{4 i (b+c) k}\right) \cos(2 a k)-2 i \left(-1+e^{2 i (b+c)
k}\right)^2 \sin(2 a k)}
{\left(-3+e^{4 i b k}+e^{4 i c
k}+e^{4 i (b+c) k}\right) \cos(2 a k)-i \left(-5+e^{4 i b k}+e^{4 i c
k}+4 e^{2 i (b+c) k}-e^{4 i (b+c) k}\right) \sin(2 a k)},
\]
\normalsize for the graph 1a, and

\small
\[
S^{(II)}_{1,1}(k)=\]
\[-\frac{2 i \sin\left((b+c) k\right) \left[\left(1-e^{4 i a k}\right)
\cos\left((b-c) k\right)+\left(1+e^{4 i a k}\right)
\cos\left((b+c) k\right)-i \left(1-e^{4 i a k}\right)
\sin\left((b+c) k\right)\right]}{1-e^{4 i a k}+\cos\left(2 (b-c) k\right)+
\left(-2+e^{4 i a k}\right) \cos\left(2 (b+c) k\right)+2 i \sin\left(2 (b+c) k\right)},
\]

\[
S^{(II)}_{1,2}(k)=\frac{2 e^{2 i a k} \sin(2 b k) \sin(2 c k)}{1-e^{4 i a
k}+\cos(2 (b-c) k)+\left(-2+e^{4 i a k}\right) \cos(2 (b+c) k)+2 i
\sin(2 (b+c) k)},
\]

\[
S^{(II)}_{2,1}(k)=S^{(II)}_{1,2}(k),
\]

\[
S^{(II)}_{2,2}(k)=\]
\[-\frac{2 i \sin\left((b+c) k\right) \left[\left(-1+e^{4 i a k}\right)
\cos\left((b-c) k\right)+\left(1+e^{4 i a k}\right) \cos\left((b+c) k\right)
-i \left(1-e^{4 i a k}\right) \sin\left((b+c) k\right)\right]}{1-e^{4 i a
k}+\cos\left(2 (b-c) k\right)+\left(-2+e^{4 i a k}\right) \cos\left(2 (b+c) k\right)+2 i \sin\left(2 (b+c)
k\right)},
\]
\normalsize
for the graph 1b.

The optical lengths of the bonds of the microwave networks are denoted by $a$, $b$, and $c$, respectively.

Let us remind that two graphs are called isopolar if their scattering matrices $S^{(I)}(k)$ and $S^{(II)}(k)$ share the same poles on the complex plane. The resonances show up as poles \cite{Kottos2003,Borthwick2007} occurring at complex wave numbers $k_l=\frac{2\pi}{c}(\nu_l -i\Delta\nu_l)$, where $\nu_l$ and $2\Delta\nu_l$ are associated with the positions and the widths of resonances, respectively. Also loses in the networks \cite{Hul04} can be effectively incorporated to the description by treating the wave number $k$  as a complex quantity with absorption-dependent imaginary part $\mathrm{Im}\Bigl[k\Bigr]=\beta \sqrt{2\pi \nu/c }$ and the real part $\mathrm{Re}\Bigl[k\Bigr] =  2\pi \nu/c $, where $\beta$ is the absorption coefficient and $c$ is here the speed of light in vacuum.

\pagebreak

\smallskip
\begin{figure}[1]

\includegraphics[width=9cm]{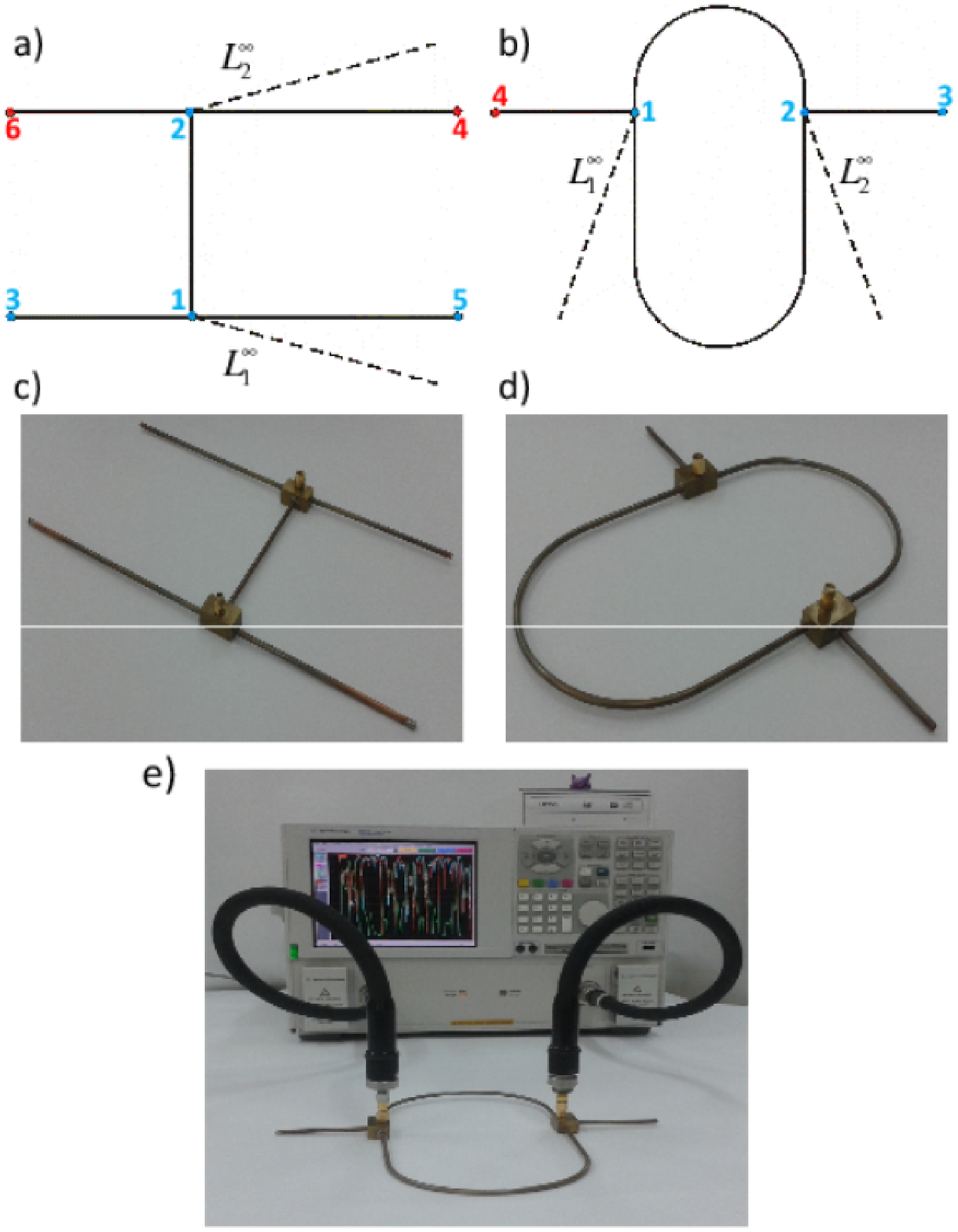}

\caption{(Color online) A pair of isoscattering quantum graphs and the pictures
of two isoscattering microwave networks are shown in the panels
({\bf a-b}) and ({\bf c-d}), respectively. Using the two
isospectral graphs, ({\bf a}) with $n=6$ vertices and ({\bf b})
with $n=4$ vertices, isoscattering quantum graphs are formed by
attaching the two infinite leads $L^{\infty}_1$ and $L^{\infty}_2$
(dashed lines). The vertices with Neumann boundary conditions
are denoted by full circles while the vertices with Dirichlet
boundary conditions by the open ones. The two isoscattering microwave
networks with $n=6$ and $n=4$ vertices which simulate quantum
graphs ({\bf a}) and ({\bf b}), respectively, are shown in the
panels ({\bf c-d}).
The connection of the microwave networks to the Vector Network Analyzer (VNA) was realized by means of the
two microwave coaxial cables (see panel {\bf e}).} \label{Fig1}
\end{figure}

\begin{figure}[2]

\includegraphics[width=9cm]{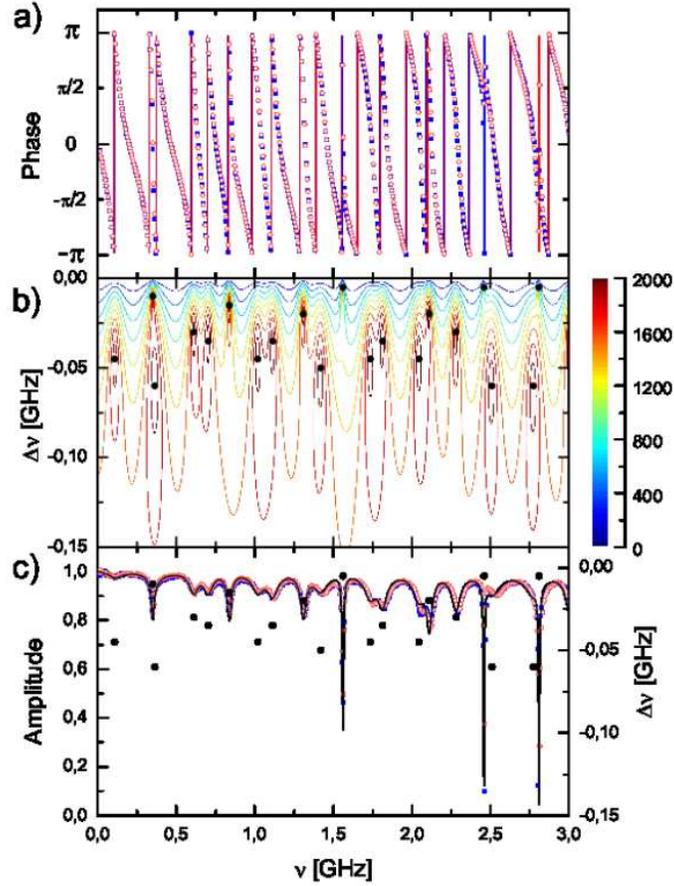}

\caption{(Color online)
({\bf a}) The amplitude of the determinant of the scattering matrix obtained
for the microwave networks with $n=4$ (red open circles) and $n=6$
(blue full squares) vertices.
The solid line shows the resonances of
the amplitude of the determinant of the theoretically evaluated scattering matrix
for the quantum graph with $n=4$ vertices. The results are presented in the frequency range $0.01-3$ GHz.
To allow the direct comparison between the positions of the theoretical poles and the experimental
and theoretical scattering resonances the poles are marked by big solid circles.
The right vertical axis of Fig.~2a shows the imaginary part $\Delta\nu$ of the poles of the graph.
({\bf b}) The contour plot shows the positions of scattering poles of
the amplitude of the determinant of the theoretically evaluated scattering matrix
for the quantum graph with $n=4$ vertices.
({\bf c}) The cumulative phase of the determinant
of the scattering matrix obtained for the microwave networks with
$n=4$ (red open circles) and $n=6$ (blue full squares) vertices.
}
\label{Fig2}
\end{figure}

\end{document}